\documentclass[%
 reprint,
superscriptaddress,
amsmath,amssymb,
aps,
pra,
]{revtex4-2}

\usepackage{graphicx}
\usepackage{dcolumn}
\usepackage{bm}
\usepackage{natbib}

\begin{document}


\title{Unexpectedly absence of Marangoni convection in an evaporating water meniscus}

\author{Tapan Kumar Pradhan}
\email{tapan.k.pradhan@gmail.com}
 \affiliation{Department of Mechanical Engineering, \\ Indian Institute of Technology Kharagpur, Kharagpur 721302, India.} 
 
\author{Pradipta Kumar Panigrahi}%
 
\affiliation{Department of Mechanical Engineering, \\ Indian Institute of Technology Kanpur, Kanpur 208016, India.}


\begin{abstract}
It is usually expected that surface tension driven flow dominates at small scales. Evaporation from the meniscus of ethanol/methanol confined in a capillary induces Marangoni convection at the meniscus which has been investigated by previous researchers. However, Marangoni convection is unexpectedly absent at an evaporating water meniscus confined inside a small micro-channel which is reported in this work. The velocity near the water meniscus is studied experimentally using confocal micro-PIV. The water meniscus is formed at the end of a micro-channel filled with water keeping the meniscus exposed to the atmosphere. The flow near the evaporating water meniscus is caused by combined effect of capillary flow to replenish water loss at the meniscus and buoyancy flow caused by evaporation induced temperature gradient. Unidirectional capillary flow dominates for small capillary size where as circulating buoyancy convection dominates in larger capillary size. The reported unexpected behavior of flow at the evaporating water meniscus will be helpful to understand the inter-facial phenomena in future studies.
\end{abstract}

\maketitle

\section{Introduction}

Evaporation induced flow is found in many micro-scale environments like droplet, meniscus, thin films etc. Evaporation from liquid meniscus has many potential application in heat pipes, micro-scale heat exchangers, porous media and many other microfluidics devices. Evaporation induced mass transfer near the meniscus can be used for coating of surfaces \cite{Dimitrov1996,Prevo2004}. Understanding the hydrodynamics of fluid interface during evaporation is very essential to study these systems. Inter-facial flow near an evaporating meniscus has been a subject matter of research by many researchers due to its potential application in these above mentioned fields. Experimental studies by \citet{Buffone2005,Dhavaleswarapu2007} show that the flow pattern near an evaporating meniscus of ethanol or methanol induces strong circulating flow near the meniscus. They observed that uneven evaporation from the curved meniscus causes temperature gradient which leads to surface tension variation along the liquid-air interface \cite{Buffone2004}. Surface tension gradient along the liquid surface induces thermocapillary convection causing circulating loops near the evaporating meniscus. For smaller channel size, two symmetric circulating loops are observed due to thermocapillary convection. When the size of the channel becomes larger, the circulation in the vertical plane becomes asymmetrical due to the effect of gravity. Numerical studies mentioned in previous literature \cite{Wang2008,Pan2013,Lan2011} show similar result where thermal Marangoni convection dominates the flow pattern in smaller scale. Such thermocapillary convection is observed both in horizontally oriented capillary \cite{Buffone2014} and vertically oriented capillary \cite{Pan2010,Song2011}. Evaporating meniscus of alcoholic solution of ethanol-water mixture also shows surface tension driven flow causing two circulating loops \cite{Cecere2014}.

\par
All the studies on evaporating meniscus mentioned in the above literature emphasize on thermocapillary convection driven by surface tension gradient. In this work, we have reported a strange behavior of evaporating water meniscus where surface tension driven flow at the interface is absent in spite of small characteristic length. The behavior of the fluid flow near the water meniscus is completely different than the reported work by previous researchers. Recently, few studies were carried out on evaporating droplet of aqueous solution where it has been reported that solutal Rayleigh convection dominates the Marangoni convection \cite{Kang2013,Pradhan2016b,Edwards2018,Li2019}. The present experimental study investigate the fluid dynamics near the evaporating water meniscus using micro-PIV technique.

\section{Experimental Details}
\begin{figure}[htb!]
\begin{center}
\includegraphics[width=0.48\textwidth]{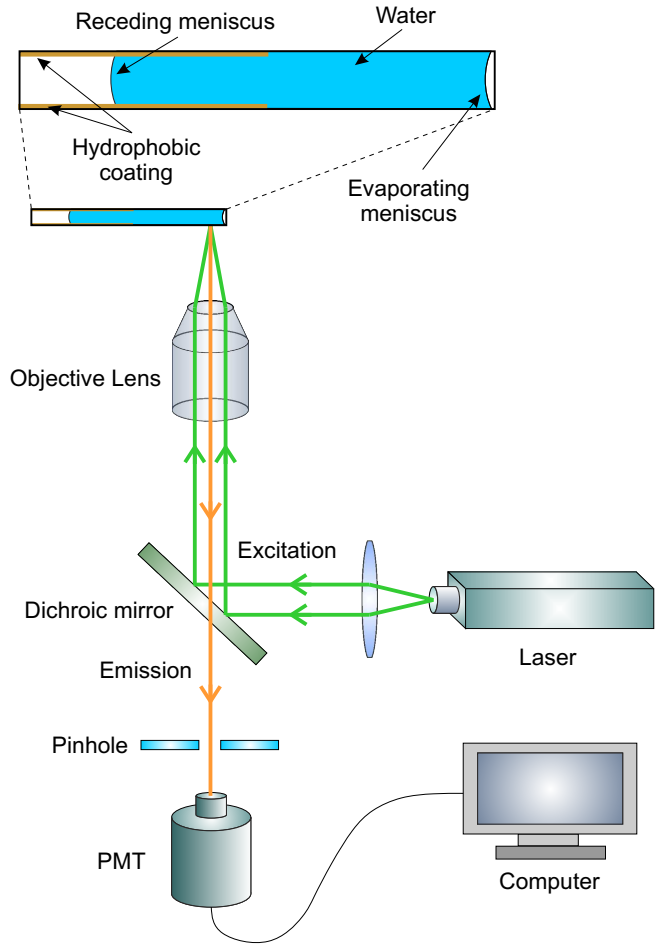}
\caption{\label{fig:exp}Experimental arrangement for measuring velocity near the evaporating water meniscus.}
\end{center}
\end{figure}

A water meniscus was formed at the end of a square glass capillary. Two square capillaries having size \(500 \; \mu \textrm{m} \times 500 \; \mu \textrm{m}\) and \(1000 \; \mu \textrm{m} \times 1000 \; \; \mu \textrm{m}\) were taken for the study. The capillaries were filled with DI water. One of the meniscus was formed at one end of the channel. The other meniscus of water lied inside the capillary. The experimental setup for the flow measurement near the meniscus is shown in Fig. \ref{fig:exp}. A part of the channel was made hydrophobic using OTS treatment \cite{Wong2013} which is shown in the left part of the channel. The hydrophobic surface allows the left meniscus to move freely without resistance inside the channel. The right meniscus remained pinned at the channel end throughout the evaporation time. The left meniscus receded and moved towards the right to compensate the loss of water at the right meniscus. The capillaries were placed in horizontal position during the velocity measurement. The right meniscus formed a contact angle of \(64^0 \pm 5^0\) at the end of the channel and remained constant throughout the evaporation process. The contact angle of the meniscus was measured using Dropsnake \cite{Stalder2006} method. The experiment was performed in an environment having ambient temperature ($T_\infty$) of \(25^0 C\) and relative humidity ($\phi$) of \(35 \; \% \pm 5 \; \%\).

\par
Velocity measurement inside the channel was carried out by micro-PIV method. Polystyrene particles tagged with fluorescent having diameter ($d_p$) 2 $\mu$m were mixed with the water for flow tracing required for PIV measurement. Flow measurement was carried out near the right meniscus. Images required for PIV evaluation were captured using a confocal microscope. A laser source (Ar ion) of wavelength 488 nm was used for illumination of the fluorescent particles. The illuminated fluorescent particles were captured by a PMT present in the confocal microscope. Size of each image was set equal to \(1024 \times 125 \; \textrm{pixels}\). The time interval between the two images was kept at 1.5 sec. Dynamic studio V1.45 (a PIV evaluation software) was used to process the images to get velocity field using adoptive cross correlation algorithm. The interrogation area during the PIV evaluation was kept equal to \(16 \times 16 \; \textrm{pixels}\) with interrogation area overlap of \(25 \%\).  

\par
The possible error caused by random Brownian motion of the tracer particles is eliminated by averaging the results over ten velocity measurements \cite{Santiago1998,Meinhart1999}. The density ($\rho$) of water at \(25^0 C\) is equal to \(997 \; \textrm{kg/m}^3\) \cite{crc}. Viscosity ($\mu$) of water at \(25^0 C\) is equal to \(8.9 \times 10^{-4} \; \textrm{Pa} \cdot \textrm{s}\) \cite{crc}. Relaxation time of the tracer particles suspended in the fluid is calculated as, \(\tau_s = d_{p}^{2} \frac{\rho_p}{18\mu}\). The relaxation time in the present work is equal to \(2.7 \times 10^{-7} \; \textrm{sec}\) which is negligible as compared to the time pulse between two consecutive images captured for PIV measurement. Hence, the particles quickly adjust themselves to the change in flow of water without any delay. The tracer particles used in the experiment have a density of \(1050 \; \textrm{kg/m}^3\). Settling velocity of the particles is given by \(U_g=\frac{d_p^2 (\rho_p-\rho)}{18\mu}g\). The settling velocity calculated for the present experiment is equal to \(0.13 \; \mu \textrm{m/s}\). The settling velocity is very less as compared to the measured fluid velocity. Hence, the effect of settling of particles on the velocity measurement of fluid can be neglected.

\section{Results and Discussion}
Evaporation induced flow near the water meniscus is presented in this section. Evaporation only occurs from the right meniscus exposed to atmosphere. There is no evaporation from the left meniscus which is inside the micro-channel due to confinement effect as studied in our previous work \cite{Pradhan2016a}. Extended channel wall near the left meniscus hinders the evaporation from the meniscus. The flow induced near the evaporating water meniscus depends on the channel size. We have studied the flow convection for two channel sizes. One is for small capillary having cross-section \(500 \; \mu \textrm{m} \times 500 \; \mu \textrm{m}\) and other is for large capillary having cross-section \(1000 \; \mu \textrm{m} \times 1000 \; \mu \textrm{m}\). The results of the both channel size have been reported in two subsections.

\begin{figure}[htb!]
\begin{center}
\includegraphics[width=0.47\textwidth]{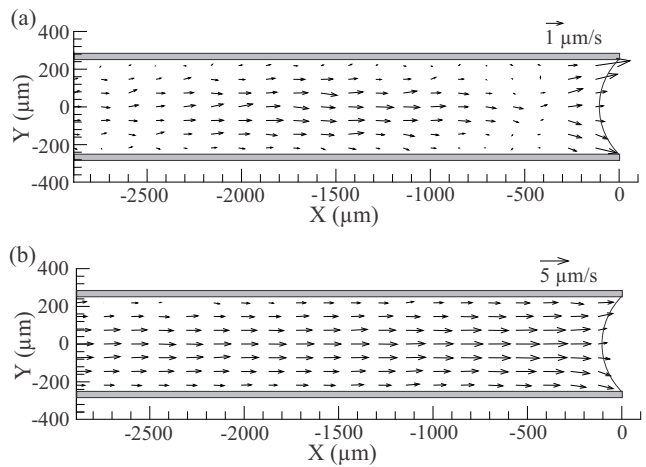}
\caption{\label{fig:small}Velocity vector field inside a small channel at (a) 100 $\mu$m and (b) 400 $\mu$m from the lower wall.}
\end{center}
\end{figure} 

\subsection{Small channel}
The flow pattern near the evaporating water meniscus for small channel case is presented in this subsection. The velocity vector field near the meniscus of small capillary is presented in Fig. \ref{fig:small} at a distance of 100 $\mu$m and 400 $\mu$m from the lower channel surface. Fluid near both the bottom and top surface moves unidirectionally towards the evaporating meniscus. All the flow field presented here is 2D velocity field. Similar velocity measurements in horizontal planes parallel to the lower channel surface were carried out at 10 X-Y planes.

\par
\begin{figure}
\begin{center}
\includegraphics[width=0.47\textwidth]{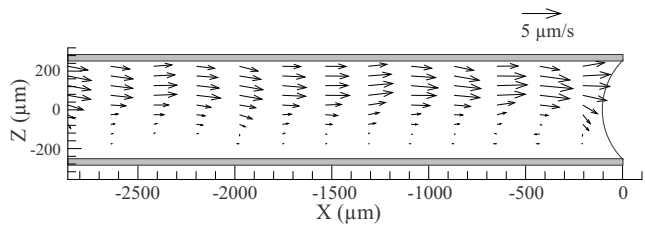}
\caption{\label{fig:smallxz} Velocity vector field perpendicular to the lower channel surface for the small channel at Y = 0.}
\end{center}
\end{figure}

Three dimensional velocity field was reconstructed using continuity equation from these two dimensional velocity measurements at different planes \cite{Pradhan2012}. The reconstructed velocity field is presented along a vertical plane at the center of the capillary as shown in Fig. \ref{fig:smallxz} to get a better picture of the flow physics. All the velocity vectors show the movement of water towards the meniscus. Evaporation from the meniscus causes loss of water at the meniscus. Water from inner region moves towards the evaporating meniscus to compensate the water loss and the flow is described as capillary flow. Similar flow is observed in evaporating droplets where fluid flows towards the pinned contact line to compensate the water loss due to evaporation as described by \citet{Deegan1997,Deegan2000}. 

\par
From the Fig. \ref{fig:small}(a) and (b), it is found that the velocity vector is higher at the edge of the evaporating meniscus. This can be explained from the evaporative flux distribution along the meniscus. The evaporative flux at the liquid-air interface of the meniscus is obtained by solving diffusion equation of water vapor in air surrounding the meniscus. Evaporation from the meniscus can be modeled as steady diffusion driven \cite{Hu2002,Sobac2011,Gelderblom2011,Carle2013,Soulie2015}. The transport of vapor from the meniscus occurs due to the difference in vapor concentration at the meniscus interface and at the ambient air. The governing equation for vapor transport is given by

\begin{equation}\label{eq:diffusion}
\nabla ^2 C=0
\end{equation}

\par
Here, $C$ denotes vapor concentration in air. The vapor concentration at the evaporating meniscus is equal to the saturated vapor concentration ($C_s$) at the ambient temperature. The ambient vapor concentration can be calculated as \(C_{\infty}=\phi C_s\). The value of $C_s$ is equal to \(2.31 \times 10^{-2} \; \textrm{kg/m}^3\) \cite{crc} and $C_{\infty}$ is equal to \(0.81 \times 10^{-2} \; \textrm{kg/m}^3\). The boundary conditions are summarized as follows,

\begin{subequations}\label{eq:boundarydiff}
\begin{align}
        C &=C_s \quad \text{at} \quad \text{meniscus surface}, \\
        C &=C_{\infty}=\phi C_{s} \quad \text{at} \quad \text{far ambient air}, \\
        \nabla C \cdot n &=0 \quad \text{at} \quad \text{channel surface}.
\end{align}
\end{subequations}

\par
The diffusion equation is solved in COMSOL Multiphysics software to get the evaporative flux from the meniscus interface. The simulation is performed in the surrounding air domain. The evaporative flux is given by 

\begin{equation}\label{eq:flux}
J=-D \nabla C \quad \text{at meniscus interface}.
\end{equation}

\begin{figure}[htb!]
\begin{center}
\includegraphics[width=0.47\textwidth]{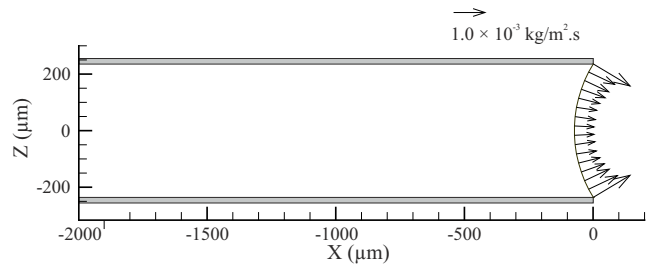}
\caption{\label{fig:evp} Evaporative flux distribution along the water meniscus.}
\end{center}
\end{figure}

\par
The evaporative flux distribution along the meniscus obtained from the simulation is presented in Fig. \ref{fig:evp}. It shows a higher evaporative flux at the edge of the meniscus as compared to the center of the meniscus. Evaporation at the meniscus drives the fluid from the inner region towards the evaporating meniscus. The evaporation induced normal velocity of the fluid at the evaporating meniscus can be given by 

\begin{equation}\label{eq:cap}
u \cdot n =\frac{J}{\rho} \quad \text{at} \quad \text{meniscus surface}.
\end{equation}

\par
The normal velocity at the evaporating meniscus is directly proportional to the evaporative flux. Hence, higher velocity is observed near the edge of the meniscus as evaporative flux is higher at this region. Maximum value of evaporative flux obtained from the simulation is equal to \(1.36 \times 10^{-3} \, \textrm{kg/m}^2\textrm{s} \). From the above equation (Eq. \ref{eq:cap}), the capillary flow velocity is found to be equal to \(1.4 \, \mu \textrm{m/s} \) which is of same order of magnitude that observed from the experiment.

\begin{figure*}
\begin{center}
\includegraphics[width=1\textwidth]{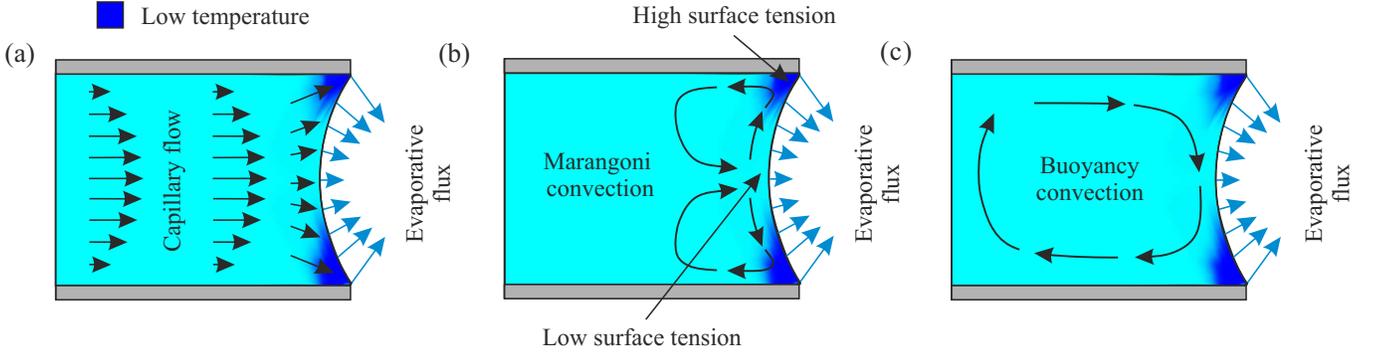}
\caption{\label{fig:depiction} Depiction of flow mechanisms near the meniscus caused by evaporation when decoupled from each other. (a) Capillary flow caused by water loss at the evaporating meniscus. (b) Thermal Marangoni convection caused by surface tension gradient at the interface. (c) Buoyancy convection due to temperature variation near the meniscus.}
\end{center}
\end{figure*}

\par
The possible flow mechanism caused by evaporation are Marangoni effect \cite{Hegseth1996,Hu2005,Karpitschka2012,Marin2016,Karpitschka2017,Darras2018,Marin2019} and buoyancy effect \cite{Pradhan2016b,Salmon2020} along with capillary flow. These mechanisms are depicted in Fig. \ref{fig:depiction} when decoupled from each other. Thermal Marangoni convection is expected to be the most dominant flow mechanism for small scale. Dominance of surface tension force can be estimated from the capillary length, \(l_c= \sqrt{\frac{\sigma}{\rho g}}\). The value of surface tension ($\sigma$) is equal to \(7.2 \times 10^{-2} \, \textrm{N/m}\) \cite{crc}.  The value of capillary length for the present case is equal to 2.7 mm. The characteristic dimension of our system (0.5 mm and 1 mm) is much less than the capillary length. Hence, surface tension effect is expected to dominate over other effects. Higher evaporative flux at the corner causes less temperature leading to high surface tension. Fluid from the low surface tension region at the center of the meniscus should flow to the region of high surface tension creating two circulating loops as presented in Fig. \ref{fig:depiction}(b). 

\par
The flow driven by surface tension gradient can be characterized by Marangoni number as, \(\mathrm{Ma}=\frac{(-d\sigma/dT)\Delta T D}{\mu \alpha}\). Temperature boundary condition  at the evaporating interface can be given by, \(-k\frac{\partial T}{\partial n}=h_{fg} J\), where $h_{fg}$ is the enthalpy of vaporization, $k$ is the thermal conductivity of water, $\alpha$ is the thermal diffusivity of water and \(\frac{\partial T}{\partial n}\) is the normal temperature gradient at the liquid-air interface. The value of \(\Delta T\) can be obtained from the scaling analysis of the temperature boundary condition. Marangoni number can be rewritten in terms of evaporative flux ($J$) as,  \(\mathrm{Ma}=\frac{(-d\sigma/dT)D^2h_{fg}J}{k\mu \alpha}\). The values of \(d\sigma/dT\), $h_{fg}$, $k$ and  $\alpha$ are equal to \(1.51 \times 10^{-4} \, \textrm{N/mK}\), \(2.44 \times 10^{6} \, \textrm{J/kg}\), \(6.06 \times 10^{-1} \, \textrm{W/mK}\) and \(1.45 \times 10^{-7} \, \textrm{m}^2/\textrm{s}\) respectively \cite{crc}. The value of Marangoni number is equal to 1602 which is much more than the critical Marangoni number, 80 (\cite{Schatz1995}). Such high Marangoni number may cause strong Marangoni convection. Unexpectedly such type of flow pattern is not observed in case of water meniscus in spite of small length scale and high Marangoni number. However, strong thermocapillary circulating flow is observed in evaporating methanol and ethanol meniscus \cite{Buffone2005,Dhavaleswarapu2007} caused by uneven  evaporation.

\par
Another flow mechanism is buoyancy effect caused by temperature gradient owing to the evaporation at the meniscus. Evaporation from the meniscus lowers the temperature which increases the density. Hence, fluid at the evaporating interface moves in downward direction creating a circulating loop as depicted in Fig. \ref{fig:depiction}(c). The flow observed in Fig. \ref{fig:smallxz} is not symmetric like capillary flow as depicted in Fig. \ref{fig:depiction}(a). The upper half shows higher flow velocity as compared to lower half though all the velocity vectors show unidirectional flow towards the evaporating meniscus. It can be explained as follows. The buoyancy effect (Fig. \ref{fig:depiction}(c)) accelerates the flow in the upper half as it has same direction as that of capillary flow. However, the buoyancy effect opposes the capillary flow in the lower half due to opposite direction. 

\begin{figure}[htb!]
\begin{center}
\includegraphics[width=0.47\textwidth]{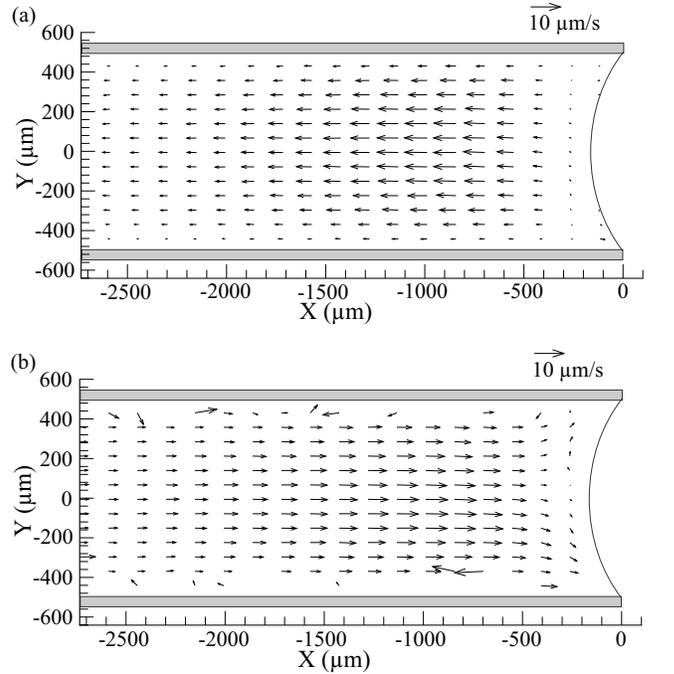}
\caption{\label{fig:large}Velocity vector field inside a large channel at (a) 100 $\mu$m and (b) 900 $\mu$m from the lower wall.}
\end{center}
\end{figure}

\subsection{Large channel}
Flow pattern inside the large channel case having size \(1000 \; \mu \textrm{m} \times 1000 \; \mu \textrm{m}\) is presented in this subsection. The velocity vector fields at 100 $\mu$m and  900 $\mu$m from the lower surface of capillary are presented in Fig. \ref{fig:large}.  Velocity vector field near the lower wall (Fig. \ref{fig:large}(a)) shows fluid flowing away from the evaporating meniscus contrary to the small channel case. Velocity vector near the upper wall (Fig. \ref{fig:large}(b)) shows fluid flow towards the evaporating meniscus. The strength of the flow for the large channel case is higher as compared to the small channel case. All these velocity vector fields in  Fig. \ref{fig:large} are 2D velocity measurements.

\par
\begin{figure}[htb!]
\begin{center}
\includegraphics[width=0.47\textwidth]{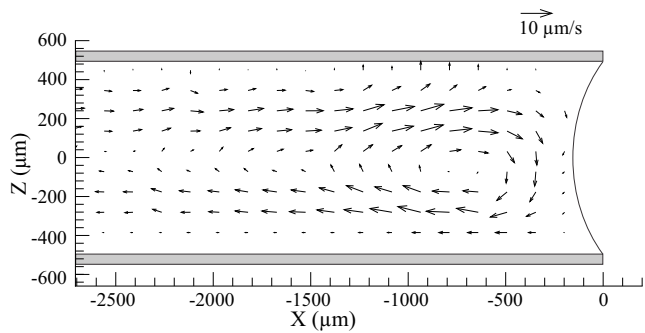}
\caption{\label{fig:largexz}Velocity vector field in a vertical plane for the large channel at Y=0.}
\end{center}
\end{figure}

Similar 2D measurements are carried out at 20 different vertical locations. Three dimensional velocity field is reconstructed from these 2D velocity measurements using continuity equation as mentioned in the previous subsection. The reconstructed velocity field in a vertical plane for the large capillary is presented in Fig. \ref{fig:largexz}. The fluid near the evaporating meniscus shows a downward movement causing circulating flow inside the channel. But only one circulating loop is observed unlike two circulating loops in case of evaporating methanol and ethanol meniscus. 

\par
Compensating flow only causes unidirectional flow which is observed for the small channel case. Here, unidirectional flow is suppressed by the recirculating flow observed in  Fig. \ref{fig:largexz}. Surface tension driven Marangoni convection is absent in the evaporating water meniscus as mentioned in the previous subsection. Hence, the effect of Marangoni convection and capillary flow can be ruled out. Hence, only thermal Rayleigh convection is left which is the cause of the fluid flow near the evaporating water meniscus. Evaporation from the meniscus reduces temperature at the meniscus causing temperature gradient near the evaporating meniscus. Temperature gradient creates density gradient causing buoyancy driven flow near the meniscus. For small capillary dimension, buoyancy force is not significant. Hence, no circulating flow is observed in the small channel case. For larger capillary dimension, the gravitational effect become dominant causing buoyancy driven natural flow.

\section{Conclusion}
We experimentally investigated the fluid flow near an evaporating water meniscus formed at the end of a capillary. The fluid flow near the meniscus was measured using confocal micro-PIV technique. The flow near the evaporating water meniscus occurs due to the combined effect of two mechanisms. First one is capillary flow towards the evaporating meniscus to replenish the water loss due to evaporation. Second one is due to buoyancy driven Rayleigh convection caused by temperature gradient as a result of evaporation. The nature of the flow pattern depends on the channel size. Fluid inside smaller capillary shows unidirectional flow towards the evaporating meniscus caused by capillary flow. When the capillary size increases, thermal buoyancy effect caused by evaporative cooling near the meniscus becomes significant. One circulating loop is observed inside the channel due the thermal buoyancy effect. The fluid near the evaporating meniscus moves downward direction along the evaporating meniscus due to buoyancy effect. Such type of flow is contrary to the flow pattern observed near the evaporating meniscus of ethanol and methanol. No surface tension driven Marangoni convection is observed near the water meniscus unlike evaporating ethanol and methanol meniscus where two circulating loops are observed due to thermal Marangoni convection. The reason for the absence of Marangoni convection is still not known. It needs further investigation at the interface to understand the phenomena.

\begin{acknowledgements}
Both the authors acknowledge the Department of Science and Technology, Government of India for the financial support and IIT Kanpur for providing facilities for the experiment. The work has been performed at the Microfluidics and sensor laboratory, IIT Kanpur.
\end{acknowledgements}

\nocite{*}
\bibliography{reference}

\end{document}